\documentclass[12pt,a4paper]{article}
\usepackage[margin=2.5cm]{geometry}
\usepackage{graphicx}
\usepackage{color}
\usepackage{amsmath,amssymb,amsfonts}
\usepackage{algorithm}
\usepackage{algorithmic}
\usepackage{float}
\usepackage[section]{placeins}
\usepackage[T1]{fontenc}
\usepackage{hyperref}
\usepackage{quantikz}
\usepackage{rotating}
\usepackage{braket}

\newlength\myindent
\newcommand\bindent{%
  \begingroup
  \setlength{\itemindent}{\myindent}
  \addtolength{\algorithmicindent}{\myindent}
}
\newcommand\eindent{\endgroup}

\begin{document}

\title{Efficient Quantum Oracle for Solving Bilinear Diophantine Equations on Digital Quantum Computers}

\author{S.~Whitlock$^{1,*}$ and T.~D.~Kieu$^2$ \\[6pt]
\small $^1$European Center for Quantum Sciences (CESQ-ISIS, UMR 7006),\\
\small University of Strasbourg and CNRS \\[2pt]
\small $^2$Centre for Quantum Technology Theory and Optical Sciences Centre,\\
\small Swinburne University of Technology, Victoria, Australia \\[4pt]
\small $^*$\texttt{whitlock@unistra.fr}}

\date{}

\maketitle


\vspace{10pt}

\begin{abstract}
We present a concrete oracle construction for bilinear Diophantine equations of the form $f(x,y) = Axy + Bx + Cy + D$, together with its application as a scalable, hardware-agnostic benchmark for digital quantum computers. The oracle can be used in a Grover search algorithm in two variants suitable for both noisy-intermediate scale quantum devices and early fault-tolerant quantum processors. Applied to integer factoring via a residue-class encoding, the circuit requires $2n-5$ qubits or fewer to factor an $n$-bit biprime $N = pq$; for $N = 143$ requiring as few as 7 qubits and 135 two-qubit gates compared to 19 qubits and 51\,048 two-qubit gates for a qubit-efficient variant of Shor's algorithm. Large-scale simulations confirm a success probability approaching 100\% for $>$800 randomly selected biprimes with $5 \leq n \leq 35$. The circuit family provides a scalable, deterministically convergent and easily verifiable benchmark in a range accessible to near term quantum hardware.
\end{abstract}

\section{Introduction and context}
Quantum computing represents a fundamental shift in how we process information which promises to solve hard computational problems out of the reach of even the most powerful classical computers. A landmark result is Shor's quantum algorithm~\cite{Shor1994}, discovered in 1994, demonstrating that quantum computers can factor composite integers with a superpolynomial speedup over the best known classical algorithms. Soon after, in 1996 Grover proposed a quantum algorithm to find a given element in an unordered database of $N$ elements in $\mathcal{O}(\sqrt{N})$ queries compared to $\mathcal{O}(N)$ required classically~\cite{Grover1996}.

Assessing the capabilities of quantum computers in a meaningful, reproducible way is an ongoing challenge. Synthetic benchmarks such as cross-entropy benchmarking~\cite{Arute2019}, quantum volume~\cite{Cross2019} and mirror circuits~\cite{Proctor2022} quantify performance without connecting to specific computational tasks. By contrast, application-oriented benchmarks such as Q-score~\cite{Martiel2021}, the MQT benchmark suite~\cite{Quetschlich2023}, and Ref.~\cite{Aboumrad2026} bring circuits closer to real problems but lack a single circuit family whose complexity scales smoothly from near-term to fault-tolerant hardware with classically verifiable outputs. Shor's algorithm comes closest to this ideal: it solves a practically relevant problem with an easily verified output, and has been demonstrated on NMR, photonic, trapped ion and superconducting platforms~\cite{Vandersypen2001,Lu2007,Lanyon2007,Lucero2012,Martin2012,Monz2016,Amico2019}. However, the resource requirements are prohibitive, and today $N = 15$ is the largest biprime factored using Shor's algorithm on a quantum device without prior knowledge of the factors embedded in the circuit~\cite{Smolin2013,Mosca2018,Gidney2021}.

In this paper we present a concrete oracle construction for bilinear Diophantine constraints $f(x,y) = M$, where $f(x,y) = Axy + Bx + Cy + D$ with integer coefficients, which can be thought of as a general integer factoring problem. By embedding the oracle in a Grover search we arrive at a benchmark algorithm for quantum computers that involves factoring $n$-bit biprimes $N = pq$ using $2n-5$ qubits or fewer. We describe two circuit variants: \emph{Grover-H}, which performs the multiply-add arithmetic in the Hadamard basis using controlled-phase gates and ancilla-free phase synthesis, suited to near-term devices; and \emph{Grover-AND}, which evaluates the zero predicate using Toffoli gates and has an exact, fixed T-gate count per oracle call, making it the preferred variant for fault-tolerant implementations. The oracle is not specific to Grover search and can be embedded in other quantum optimization frameworks; replacing the diffusion operator with an adiabatic schedule yields a digital adiabatic variant described in the Supplementary Information. We note that the present Grover based approach provides no speedup over classical factoring, but does provide a valuable benchmark for digital quantum computers: the output is a binary pass/fail classically verifiable, the circuit is fully determined by $N$ with no instance-specific tuning (in contrast to variational algorithms), and the difficulty scales smoothly with $n$. We confirm convergence and scaling through large-scale simulations on $>$800 randomly selected biprimes with $5 \leq n \leq 35$.

\section{Quantum factoring algorithm using Grover search}\label{sec:algorithm}

\begin{figure}[t]
\centering
\includegraphics[width=0.95\linewidth]{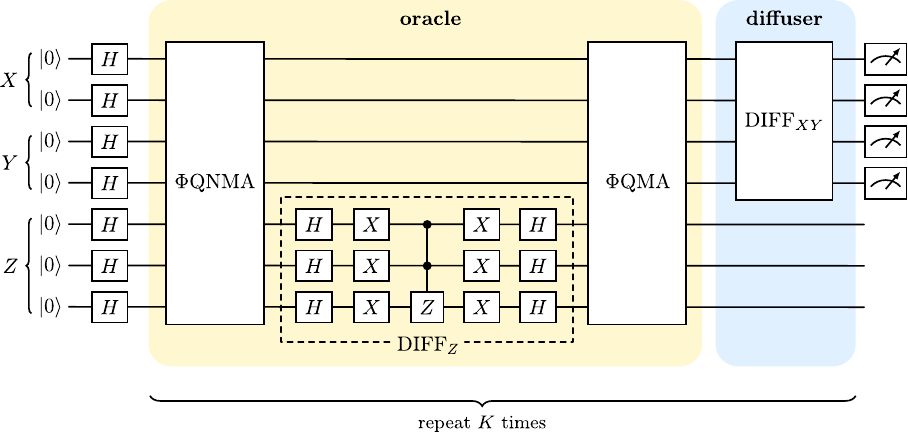}
\caption{Quantum algorithm for factoring a biprime $N$ using Grover search (Grover-H variant). The $X$ and $Y$ registers are initialised in the uniform state by a single Hadamard layer applied once before the loop. The $Z$ register is initialised in $\ket{+}^{\otimes n_z}$ by a Hadamard layer applied once before the loop, replacing the state preparation and QFT of the unoptimised circuit. Each step of the loop applies an oracle ($\Phi$QNMA, a Hadamard layer, $X$ gates, a multicontrolled-$Z$ gate, $X$ gates, and a Hadamard layer --- together forming $\mathrm{DIFF}_Z$ --- followed by $\Phi$QMA) and the diffuser $\mathrm{DIFF}_{XY}$. A final Hadamard layer on $Z$ is applied once after the loop. The output is obtained by measuring the $X$, $Y$ registers in the computational basis.}
\label{fig:circuit1}
\end{figure}

Grover's search algorithm efficiently locates a marked item in an unstructured database with $2^n$ entries using $\mathcal{O}(2^{n/2})$ queries, providing a quadratic speedup over classical searching. It can also be used to find the prime factors $p, q$ of a $n$-bit integer $N=pq$. To illustrate, consider two quantum registers $X$, $Y$, with computational basis states $\ket{x},\ket{y}$, both initialized in the uniform state. One then applies a loop, consisting in each step of an oracle operator which applies a $\pi$ phase shift $|x,y\rangle \rightarrow -|x,y\rangle$ iff $xy = N$, where $x,y$ is the integer representation of $\ket{x},\ket{y}$. A concrete example of such an oracle operation built from elementary quantum circuits is shown in Fig.~\ref{fig:circuit1}. This is followed by the Grover diffuser operation~\cite{Nielsen2010} which increases the amplitude in the marked state. The probability in the solution state increases in each step, approaching $100\%$ after $\sim 2^{n/2}$ steps, such that the factors $p, q$ can be directly read out by simply measuring $X$, $Y$ in the computational basis. The approach generalises straightforwardly to general quadratic Diophantine equations in two or more variables.

Algorithm~\ref{alg:1} and Fig.~\ref{fig:circuit1} present the Grover-H factoring algorithm using $2n-5$ qubits. The residue-class encoding reduces the search space: primes and biprimes with $p, q > 3$ satisfy $N = 6(M+1) + S$, $p = 6(a+1) + s$, $q=6(b+1) + sS$ with $M, a, b \geq 0$ and $S,s = \pm 1$, so factoring reduces to finding $a, b$ satisfying $f(a,b) = 0$ where $f(a,b) = 6(a+1)(b+1) + s(b+1) + sS(a+1) - 1 - M$ and $M = (N-S)/6-1$. Convergence of the algorithm is guaranteed within $K_\mathrm{opt}=\lfloor (\pi/4) 2^{(n_x+n_y)/2}\rfloor$ steps (or $K_\mathrm{opt}=\lfloor (\pi/4) 2^{(n_x+n_y)/2}/\sqrt{2}\rfloor$ when two solutions fit the register sizes, e.g.\ balanced factors).

The oracle can be decomposed in terms of several circuit primitives as shown in Fig.~\ref{fig:circuit1} (see Supplementary Information for a full decomposition in terms of elementary gates). A key operation is the quantum multiply-and-add gate (QMA; not to be confused with the complexity class Quantum Merlin Arthur) and its inverse (QNMA), which are similar to the fused multiply-add and fused negative multiply-add instructions in modern CPUs. These operations act on all three registers to realize the unitary transformation $|x,y\rangle|z\rangle \rightarrow |x,y\rangle|z\pm (Axy + Bx + Cy + D) \mod 2^{n_z}\rangle$ where $A,B,C,D$ are integer parameters. This can be realized with $\mathcal{O}(n^3)$ gates in the Fourier domain as a sequence of one-qubit phase gates and two- and three-qubit controlled phase gates acting on the $Z$ register, preceded by a quantum Fourier transform and followed by an inverse quantum Fourier transform. The QNMA, QMA operations envelop a (inverted) multicontrolled Z gate which applies a $\pi$ phase shift to the $|z=0\rangle$ basis state. This is implemented via phase kickback ($H \cdot C^{n_z-1}X \cdot H$) using the Barenco dirty-ancilla ladder~\cite{Barenco1995}, with the $X, Y$ registers supplying the $n_z - 3$ required dirty ancilla qubits, giving a cost of approximately $4(n_z - 3)$ Toffoli gates.

\label{par:grover-h-opt}The two variants share the same overall structure but differ in how the oracle is implemented. Grover-H initialises $Z$ in $\ket{+}^{\otimes n_z}$ and performs the arithmetic directly in the Hadamard basis using controlled-phase gates, eliminating the QFT from the circuit entirely. Grover-AND instead keeps $Z$ in $\ket{0}^{\otimes n_z}$ and evaluates the zero predicate $f(x,y)=0$ in the computational basis using Toffoli gates (Supplementary Information); it requires two additional clean ancilla qubits beyond the $(X, Y, Z)$ registers (one to flag the zero condition and one scratch qubit for the AND-merged control) giving a qubit count of $2n-3$ in the worst case. The two Hadamard layers flanking the arithmetic, together with the $X$ gates and multicontrolled-$Z$ gate, implement $\mathrm{DIFF}_Z$: a phase flip on $\ket{+}^{\otimes n_z}$ that marks the solution state $f(x,y)=0$. In terms of elementary gates, the entire algorithm requires $\mathcal{O}(n^3 2^{n/2})$ gates; the Supplementary Information includes a full breakdown.

The maximum number of qubits required for the algorithm can be constrained depending on the number of bits in $N$ as
$$n_x = \left\lfloor \frac{n}{2}-2-d\right\rfloor, \qquad n_y=\left\lceil \frac{n}{2}-2+d\right\rceil$$
where $d$ is the bit distance $d = (n_y-n_x)/2$. In many cases of practical relevance such as the RSA cryptosystem, the bit distance is close to zero. Otherwise the algorithm must be run for different values of $d$ in the range $0\leq d \leq \lfloor \frac{n}{2}-2\rfloor$ until a solution is found. Since $n_x+n_y \leq n-4$ (residue-class encoding) and $n_z = n_x+n_y+3$ (to prevent overflow), the algorithm requires a total of $2n-5$ qubits in the worst case. For balanced factors satisfying $\lceil\log_2 p\rceil = \lceil\log_2 q\rceil$, the qubit count can be as low as $2n-9$.

\begin{algorithm}[ht]
\small
\caption{Factorize($N$) using Grover search (Grover-H variant)}
\begin{algorithmic} \label{alg:1}
    \STATE \textbf{Input:}
    \bindent
        \STATE $\bullet$ The integer $N$ to be factored and the number of iteration steps $K$. \\
        \qquad\quad For convenience we define $M=(N-S)/6-1$ and $S=\frac{3}{2} - \frac{1}{2}(N\mathrm{~mod~} 6)$
        \STATE $\bullet$ Three registers $X, Y, Z$ containing $n_x$, $n_y$ and $n_z = n_x+n_y+3$ qubits respectively.
    \eindent
    \STATE \textbf{Output:}
    \bindent
        \STATE $\bullet$ Two bitstrings $x=a, y=b$ satisfying $N=(6a+6+s)(6b+6+sS)$ with $s=\pm 1$
    \eindent
    \STATE \textbf{Procedure:}
    \bindent
        \STATE \textbf{Step 0.} Check that $N$ is not a multiple of $2$ or $3$. Otherwise return the trivial factors.
        \STATE \textbf{Step 1.} Initialize registers\\
        \qquad\qquad$\bullet$ Prepare $X$ and $Y$ registers in the uniform state, i.e., $\frac{1}{\sqrt{2^{n_x+n_y}}}\left (\ket{0}+\ket{1}\right )^{\otimes (n_x+n_y)}$\\
        \qquad\qquad$\bullet$ Prepare $Z$ register in the uniform state $\ket{+}^{\otimes n_z}$ (Grover-H; see Supplementary Information for Grover-AND)
        \STATE \textbf{Step 2.} Choose $s=1$
        \FOR{ $1 \leq k \leq K$}
        \bindent
        \STATE \textbf{Step 3.} Apply oracle: mark all $(x,y)$ satisfying $f(x,y)=0$ with a $\pi$ phase shift\\
        \qquad\qquad$\bullet$ Apply $\Phi$QNMA: each $\ket{z}$ component of $Z$ acquires phase $e^{-2\pi i f(x,y)z/2^{n_z}}$,\\
        \qquad\qquad\phantom{$\bullet$} when $f(x,y)=0$ all phases vanish and $Z = \ket{+}^{\otimes n_z}$\\
        \qquad\qquad$\bullet$ Apply $\mathrm{DIFF}_Z = H^{\otimes n_z} X^{\otimes n_z} \cdot \mathrm{C}^{n_z}\mathrm{Z} \cdot X^{\otimes n_z} H^{\otimes n_z}$;\\
        \qquad\qquad\phantom{$\bullet$} flips the sign of the solution state\\
        \qquad\qquad$\bullet$ Apply $\Phi$QMA: uncompute, restoring $Z$ to $\ket{+}^{\otimes n_z}$
        \STATE \textbf{Step 4.} Apply Grover's diffusion operator $\mathrm{DIFF}_{XY}$
        \eindent
        \ENDFOR
    \STATE \textbf{Step 5.} Measure the $X$, $Y$ registers in the $0,1$ basis. Calculate $p=6x+s, q=6y+sS$. \\
    \qquad\qquad\quad\quad If the computed $p,q$ do not satisfy $pq=N$, restart with $s=-1$ in Step~2.
    \eindent
\end{algorithmic}
\end{algorithm}

\section{Circuit variants and comparison with Shor's algorithm}

We compare gate resources for Grover-H and Grover-AND against three variants of Shor's algorithm for $N = 143 = 11 \times 13$ ($n = 8$, $K = 1$; Table~\ref{tab:resources}). Grover-H requires far fewer multiqubit gates than Grover-AND for small $n$, but controlled-phase gates with non-$\pi/4$ angles cannot be expressed exactly in the Clifford+T gate set required by fault-tolerant hardware and must be approximated by T-gate sequences using the Ross--Selinger algorithm~\cite{RossSelinger2016}. Grover-AND evaluates the zero predicate $\ket{z=0}$ using Toffoli gates in the computational basis; it has an exact T-gate count of $7n_\mathrm{CCX}$ per oracle call, making it the preferred variant for fault-tolerant implementations. The three Shor variants differ in their phase estimation register size: the full implementation uses $4n+2$ qubits~\cite{Nielsen2010}, the reduced version uses $3n+2$~\cite{Nielsen2010}, and the iterative (Beauregard) variant uses $2n+3$ qubits via mid-circuit measurement and classical feedforward~\cite{Beauregard2002}.

\begin{table}[ht]
\centering
\begin{tabular}{lrrrrr}
\hline
 & & \multicolumn{3}{c}{Elementary} & C+T \\
Algorithm & Qubits & 1-qubit & 2-qubit & 3-qubit & T-gates \\
\hline
Grover-H              &  7 &     52 &     21 &    16 &   1\,118 \\
Grover-AND            &  9 &     39 &     21 &   222 &   1\,344 \\
\hline
Shor full ($4n+2$)      & 34 & 12\,371 & 54\,000 & 5\,957 & 495\,106 \\
Shor reduced ($3n+2$)   & 26 &  6\,195 & 27\,016 & 2\,881 & 233\,054 \\
Shor iterative ($2n+3$) & 19 &  6\,239 & 26\,976 & 2\,881 & 233\,054 \\
\hline
\end{tabular}
\caption{Gate resources for Grover-H ($7$ qubits), Grover-AND ($9$ qubits), and three variants of Shor's algorithm for $N = 143 = 11\times 13$ ($n = 8$, $K = 1$). Elementary gate counts use the 1-, 2-, and 3-qubit gates of the circuit construction without further decomposition; Clifford+T (C+T) counts decompose all multiqubit gates into T-gates. Grover-AND uses exact Toffoli decomposition; Grover-H uses Ross--Selinger synthesis~\cite{RossSelinger2016} at $\varepsilon = 0.1$. The Shor iterative variant uses mid-circuit measurements; T-gate counts are obtained by decomposing all unitary gates and leaving measurements in place.}
\label{tab:resources}
\end{table}

For the smallest non-trivial instance $N=143$ ($n = 8$), Grover-H requires $7$ qubits compared to $34$, $26$, and $19$ for the three Shor variants. Before decomposition, Grover-H requires only 16 three-qubit gates compared to 222 for Grover-AND and 2\,881 for Shor iterative, which would be particularly costly on near-term hardware as they typically decompose into several two-qubit gates with accumulated error. Since the native gate sets differ (Grover-H and Shor use controlled-phase (CCP) gates while Grover-AND uses Toffoli gates) we also compare resources in a Clifford+T decomposition. Grover-H requires 135 two-qubit gates and 1\,118 T-gates compared to 51\,048 and 233\,054 for Shor iterative, a reduction of $380\times$ and $210\times$ respectively at $n=8$. Grover-AND has a higher T-gate count (1\,344) than Grover-H at this size due to the Toffoli decomposition, but it is exact. For larger $n$, Grover-H is less competitive in the fault-tolerant regime: the $\mathcal{O}(n^3)$ phase rotations per oracle call accumulate synthesis error, so each rotation must be approximated to tighter precision, increasing the T-gate cost per rotation (Supplementary Information). Grover-AND avoids this entirely: its T-gate count scales predictably as $7n_\mathrm{CCX}$ with no precision penalty. Although Grover-H carries a much smaller prefactor, the $\mathcal{O}(n^3 2^{n/2})$ growth of the Grover iteration count dominates over the $\mathcal{O}(n^3 \log n)$ growth of Shor iterative; the two algorithms reach comparable total gate counts at $n \gtrsim 20$.

\section{Proposed benchmark protocol}

The circuit family presented here is immediately deployable as a benchmark across NISQ devices, quantum emulators, and early fault-tolerant processors on equal footing: the same circuit family, fully determined by $N$ and $K$ with no instance-specific tuning, runs on all platforms. The smallest instance, $N = 143$ ($n = 8$, 7 qubits, 135 two-qubit gates in the Clifford+T decomposition), should be within reach of any digital quantum computer with $\geq 8$ qubits and two-qubit gate fidelities above $\sim 0.99$. There are 59 biprimes with $n \leq 9$ (both prime factors $> 3$, as required by the residue-class encoding), all requiring at most 11 qubits --- a range accessible to current NISQ hardware.

The primary benchmark metric is the success probability $P_\mathrm{succ}$ above the random-sampling baseline $1/2^{n_x+n_y}$, reported at the largest instance size $N$ for which the device achieves $P_\mathrm{succ}$ significantly above baseline. If a benchmark attempt requires additional trials (e.g.\ both values of $s$, or multiple register-size choices), the total number of trials should be reported alongside $P_\mathrm{succ}$. The output is binary pass/fail --- a shot succeeds if the measured bitstrings decode to prime factors $p = 6a + s$, $q = 6b + sS$ satisfying $pq = N$, verified by a cheap multiplication check --- so there is no ambiguity in scoring. Any necessary circuit compilation and further gate-level optimisations are welcome provided no knowledge of the factors is embedded in the circuit preparation.

Both Grover-H and Grover-AND circuits are provided for $n = 8$--$20$ in OpenQASM 2.0 format covering 39 biprimes in total (covering balanced, unbalanced and highly-unbalanced factors), targeting NISQ and early fault-tolerant hardware respectively with the same problem instances. The provided circuits select the correct sign $s$ and register sizes. Additional problem instances for arbitrary $N$ can be generated using the accompanying code (see data availability statement). At $n \approx 20$ the Grover circuit depth, growing as $\mathcal{O}(2^{n/2})$, begins to exceed the resource requirements of Shor's algorithm; reaching and demonstrating this crossover regime would itself be a meaningful milestone, marking the scale at which practical quantum factoring becomes competitive.

\section{Numerical simulations}

We evaluate the Grover-H algorithm using a high performance quantum circuit simulator by QPerfect~\cite{QPerfect}. We use the exact state vector simulator which exploits highly optimized SIMD instructions to efficiently simulate deep quantum circuits. To study the largest possible range of biprimes we have also implemented a compiled version of the algorithm which applies the oracle directly to a state vector without the need for the auxiliary $Z$ register~\cite{Willsch2023}. This makes it possible to simulate the algorithm exactly for problem sizes up to $n\approx 35$ on a desktop computer, corresponding to $\approx 65$ qubits. We verified that the compiled algorithm yields the same results as the decomposed version for problem sizes that can be simulated with the decomposed algorithm.

\begin{figure}[H]
    \centering
    \includegraphics[width=0.65\linewidth]{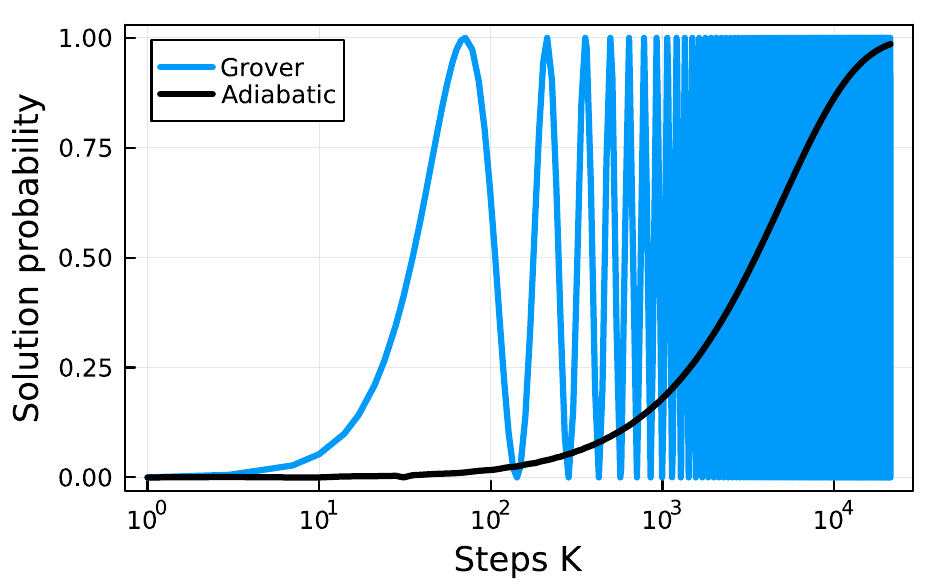}
    \caption{Success probability for $N=101911= 223\times 457$ ($n=17$, $n_x+n_y=13$) as a function of iteration steps $K$. Grover-H (blue) reaches $P \approx 1$ after $K_\mathrm{opt} = 71$ steps. For comparison, the digital adiabatic variant (black, linear schedule, $\epsilon=0.45$) is shown; with this schedule it reaches $P>0.9$ only after $\gtrsim 10^4$ steps, though convergence depends on the schedule and is not guaranteed to be optimal. The digital adiabatic variant is discussed further in the Supplementary Information.}
    \label{fig:comparison}
\end{figure}

Figure~\ref{fig:comparison} shows simulation results for the Grover-H factorization algorithm for $N = 101911 = 223\times 457$ ($n=17$ corresponding to a simulation with $n_x+n_y = 13$) as a function of the number of iteration steps. The algorithm reaches a probability $P\approx 1$ in the solution state in just $K_\mathrm{opt} = 71$ steps and afterwards undergoes periodic oscillations. Qualitatively similar behaviour is seen for all tested biprimes. Fig.~\ref{fig:comparison} also shows, for reference, results from a digital adiabatic variant of the same oracle using a linear annealing schedule which is seen to converge much more slowly.

\begin{figure}[H]
    \centering
    \includegraphics[width=0.9\linewidth]{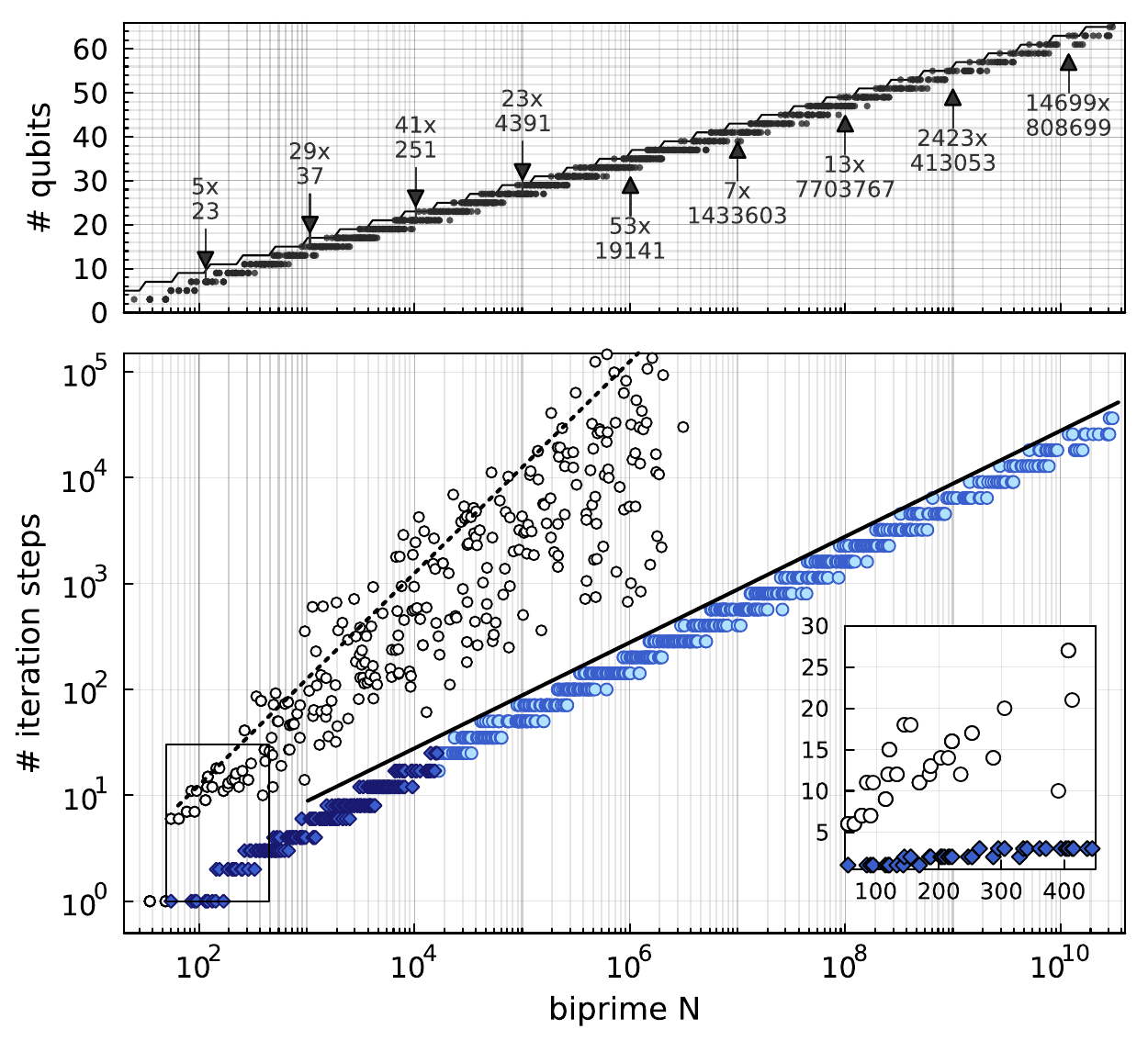}
    \caption{Simulation results and required quantum resources for the Grover-H factoring algorithm for randomly selected biprimes $N$ from 25 to $\approx 2^{35}$. (a) Total number of qubits versus problem size $N$. The black line depicts $2n-5$ scaling where $n$ is the number of bits in $N$; the lower envelope follows $2n-9$ for balanced factors. (b) Total number of Grover iteration steps versus problem size $N$. Dark diamonds indicate simulations of the full decomposed quantum circuit; light circles use the compiled oracle. The solid black line shows the asymptotic scaling $\pi 2^{n/2-4}$. Open circles show, for comparison, the number of steps required by the digital adiabatic variant (same oracle, linear schedule); the large scatter and higher step counts reflect the non-optimal schedule. The inset shows a zoomed-in view of the region indicated by the box.}
    \label{fig:qscaling}
\end{figure}

We have tested the Grover-H algorithm on $>800$ factoring problems randomly selected with $n\leq 35$, confirming successful factorization in all cases with $P\rightarrow 1$. Figure~\ref{fig:qscaling} summarizes the minimum required number of qubits and iteration steps for each problem. The qubit count ranges from $2n-9$ to $2n-5$ depending on the factors. Convergence is deterministic, with the required number of iteration steps scaling asymptotically as $\pi 2^{n/2-4}$. The number of gates per iteration step is dominated by the multiply and add operations scaling as $n^3$. The scale of the simulations, here obtained using a single CPU with 64 GiB of total memory, are comparable to recent results for a highly optimized version of Shor's algorithm emulated on a 2048 GPU cluster~\cite{Willsch2023}. The largest biprime we have factored on an emulator is $N=30398263859 = 7393 \times 4111763$ ($n=35$).

\section{Conclusion}
We have presented a concrete oracle construction for bilinear Diophantine equations $f(x,y) = M$, implemented in two circuit variants suited to both NISQ and early fault-tolerant quantum computers. Applied to integer factoring, the circuit requires $2n-5$ qubits in the worst case and $2n-9$ qubits in the best case. Grover's algorithm is optimal for unstructured problems~\cite{Bennett1997,Zalka1999}; however, the Grover-based algorithm provides no asymptotic speedup over classical factoring and is presented as a benchmark and oracle construction, not as a route to quantum advantage. Interestingly, our oracle construction works equally well when the QFT/IQFT operations used in standard quantum arithmetic~\cite{Ruiz2017} are replaced by simple Hadamard layers, eliminating the quantum Fourier transform from the circuit entirely. The circuit family provides a hardware benchmark with binary pass/fail output, classically verifiable answers, no instance-specific parameter tuning, and smoothly scalable complexity enabling systematic cross-platform comparison on equal footing.

In both variants the $Z$ register is physically initialised in $\ket{0}^{\otimes n_z}$, with the target $M$ absorbed into the constant $D$ of the oracle. However, initialising $Z$ in an arbitrary computational-basis state $\ket{z_0}$ shifts the marked condition from $f(x,y) = 0$ to $f(x,y) = z_0 \pmod{2^{n_z}}$, allowing different instances of the Diophantine equation to be solved without modifying the oracle.

The oracle construction extends straightforwardly beyond the bilinear case $f(x,y) = Axy + Bx + Cy + D$ to the most general inhomogeneous quadratic Diophantine equation in $m$ variables,
\begin{equation}
  \mathbf{x}^T Q\, \mathbf{x} + \boldsymbol{\omega}^T \mathbf{x} = M,
  \label{eq:quadratic-diophantine}
\end{equation}
where $Q$ is an integer matrix and $\boldsymbol{\omega}$ an integer vector, by adding one multiply-add stage per additional variable or quadratic term. Support for quadratic diagonal terms $Ux^2 + Vy^2$ is already included in the accompanying code repository. Integer factoring (without residue-class encoding) corresponds to the rank-2 symmetric bilinear case $Q = \mathbf{e}_1 \mathbf{e}_2^T + \mathbf{e}_2 \mathbf{e}_1^T$, $\boldsymbol{\omega} = \mathbf{0}$, $M = N$. When $Q$ is positive definite, determining whether Eq.~(\ref{eq:quadratic-diophantine}) admits a solution over $\mathbb{Z}^m$ is equivalent to the Closest Vector Problem (CVP) on the lattice with Gram matrix $Q$~\cite{vanEmde1981}, whose exact version is NP-hard. Unlike factoring-based cryptography, which is broken by Shor's algorithm, no efficient quantum algorithm is currently known for CVP in full generality. The oracle construction presented here thus provides a concrete, compilable circuit for a class of problems that includes both a practical near-term benchmark and connections to some of the hardest problems in computational number theory.

\vspace{10pt}
\noindent\textbf{Acknowledgements} We acknowledge valuable discussions with Johannes Schachenmayer and Guido Masella. This work has benefited from a state grant managed by the French National Research Agency under the Investments of the Future Program with the reference ANR-21-ESRE-0032 ``aQCess - Atomic Quantum Computing as a Service'', the Horizon Europe programme HORIZON-CL4-2021-DIGITAL-EMERGING-01-30 via the project ``EuRyQa - European infrastructure for Rydberg Quantum Computing'' grant agreement number 10107014 and support from the Institut Universitaire de France (IUF).

\vspace{10pt}
\noindent\textbf{Data availability} The data reported in this manuscript and Python scripts for constructing the circuit are available from https://github.com/aQCess/QuantumFactoring.

\vspace{10pt}
\noindent\textbf{Competing interests} SW is co-founder and shareholder of QPerfect.

\bibliographystyle{iopart-num}
\bibliography{main}

\end{document}

\title{Supplementary Information:\\Efficient Quantum Oracle for Solving Bilinear Diophantine Equations on Digital Quantum Computers}

\author{S.~Whitlock$^1$ and T.~D.~Kieu$^2$ \\[6pt]
\small $^1$European Center for Quantum Sciences (CESQ-ISIS, UMR 7006),\\
\small University of Strasbourg and CNRS \\[2pt]
\small $^2$Centre for Quantum Technology Theory and Optical Sciences Centre,\\
\small Swinburne University of Technology, Victoria, Australia}

\date{}

\maketitle

\setcounter{algorithm}{1}
\section*{S1. Grover-AND variant}

Algorithm~2 gives the Grover-AND variant of the factoring algorithm. It differs from Algorithm~1 of the main text in steps~1 and~3 only: no Hadamard layer is applied to $Z$, and the oracle evaluates the zero predicate using Toffoli gates in the computational basis rather than phase arithmetic in the Hadamard basis.

\begin{algorithm}[h]
\small
\caption{Factorize($N$) using Grover search (Grover-AND variant)}
\begin{algorithmic} \label{alg:2}
    \STATE \textbf{Input:}
    \bindent
        \STATE $\bullet$ The integer $N$ to be factored and the number of iteration steps $K$.\\
        \qquad\quad For convenience we define $M=(N-S)/6-1$ and $S=\frac{3}{2}-\frac{1}{2}(N\mathrm{~mod~}6)$
        \STATE $\bullet$ Three registers $X, Y, Z$ containing $n_x$, $n_y$ and $n_z = n_x+n_y+3$ qubits respectively.
    \eindent
    \STATE \textbf{Output:}
    \bindent
        \STATE $\bullet$ Two bitstrings $x=a, y=b$ satisfying $N=(6a+6+s)(6b+6+sS)$ with $s=\pm 1$
    \eindent
    \STATE \textbf{Procedure:}
    \bindent
        \STATE \textbf{Step 0.} Check that $N$ is not a multiple of $2$ or $3$. Otherwise return the trivial factors.
        \STATE \textbf{Step 1.} Initialize registers\\
        \qquad\qquad$\bullet$ Prepare $X$ and $Y$ registers in the uniform state, i.e., $\frac{1}{\sqrt{2^{n_x+n_y}}}\left(\ket{0}+\ket{1}\right)^{\otimes(n_x+n_y)}$\\
        \qquad\qquad$\bullet$ Prepare $Z$ register in $\ket{0}^{\otimes n_z}$
        \STATE \textbf{Step 2.} Choose $s=1$
        \FOR{$1 \leq k \leq K$}
        \bindent
        \STATE \textbf{Step 3.} Apply oracle\\
        \qquad\qquad$\bullet$ Apply $\Phi$QNMA: $\ket{x,y}\ket{z}\rightarrow \ket{x,y}\ket{z - f(x,y) \bmod 2^{n_z}}$ (computational basis)\\
        \qquad\qquad$\bullet$ Mark solution state: apply $X^{\otimes n_z}$, a multicontrolled-$X$ on the flag qubit, then $X^{\otimes n_z}$ ---\\
        \qquad\qquad\phantom{$\bullet$} flipping the flag iff $z=0$; apply $Z$ on the flag qubit to impart a $\pi$ phase shift\\
        \qquad\qquad$\bullet$ Apply $\Phi$QMA: uncompute, restoring $Z$
        \STATE \textbf{Step 4.} Apply Grover's diffusion operator $\mathrm{DIFF}_{XY}$
        \eindent
        \ENDFOR
    \STATE \textbf{Step 5.} Measure the $X$, $Y$ registers in the $0,1$ basis. Calculate $p=6x+s, q=6y+sS$.\\
    \qquad\qquad\quad\quad If the computed $p,q$ do not satisfy $pq=N$, restart with $s=-1$ in Step~2.
    \eindent
\end{algorithmic}
\end{algorithm}

\section*{S2. Basic Grover factoring oracle}

We describe the factoring algorithm without the residue-class encoding of Section~2 and without the Hadamard-basis optimisation, to give a self-contained starting point for implementation.

\textbf{Registers.} Three registers are used: $X$ ($n_x = \lceil n/2 \rceil$ qubits) and $Y$ ($n_y = n - n_x$ qubits) for the factor candidates, and $Z$ ($n_z = n_x + n_y + 1$ qubits) as arithmetic workspace. Qubits are numbered LSB-first (qubit~0 is the least significant bit). The QFT convention is
\begin{equation}
\mathrm{QFT}\ket{z} = \frac{1}{\sqrt{2^{n_z}}}\sum_{k=0}^{2^{n_z}-1}e^{2\pi ikz/2^{n_z}}\ket{k}.
\end{equation}

\textbf{Initialization.} Prepare $X$ and $Y$ in the uniform superposition by applying $H$ to each qubit. Prepare $Z$ in $\ket{z = N}$ by applying $X$ gates to the qubits corresponding to set bits of $N$.

\textbf{Oracle.} The oracle marks all $(x,y)$ satisfying $xy = N$ with a $\pi$ phase shift and leaves $Z$ unchanged. Define $f(x,y) = xy$ (bilinear form with $A=1$, $B=C=D=0$, target $M=N$). One oracle call consists of:
\begin{enumerate}
  \item Apply $\Phi$QNMA: $\ket{x,y,z} \rightarrow \ket{x,y,z - f(x,y) \bmod 2^{n_z}}$ (QFT on $Z$; controlled Fourier-domain additions; IQFT on $Z$).
  \item Apply a multicontrolled-$Z$ gate on all $n_z$ qubits of $Z$ (phase flip on $\ket{0}^{\otimes n_z}$).
  \item Apply $\Phi$QMA to uncompute step~1, restoring $Z$.
\end{enumerate}

\textbf{Grover step.} Apply the oracle followed by the diffuser $\mathrm{DIFF}_{XY}$ on $X \otimes Y$. Repeat for $K = \lfloor (\pi/4)\, 2^{(n_x+n_y)/2} \rfloor$ steps.

\textbf{Output.} Measure $X$ and $Y$ in the computational basis. The bitstrings give factors $p = x$, $q = y$ provided $xy = N$.

\textbf{Relation to the optimised algorithm.} The residue-class encoding of Section~2 replaces the direct search over $(x, y)$ with a search over $(a, b)$, reducing $n_x + n_y$ by approximately 4 bits for balanced factors. The Hadamard-basis optimisation additionally eliminates the $\ket{z = N}$ initialisation and the QFT/IQFT pair in the oracle by initialising $Z$ in $\ket{+}^{\otimes n_z}$, replacing both QFT layers with a single Hadamard layer.

\section*{S3. Decomposition of QMA in terms of elementary gates}

The QMA operation performs the unitary transformation
$$QMA=\sum_{x,y,z}\ket{x,y, z+axy+bx+cy+d \bmod 2^{n_z}}\bra{x,y,z}$$
where $a,b,c,d$ are integer coefficients (corresponding to $A,B,C,D$ in the main text) and $x,y,z$ refer to the basis states of the $X, Y, Z$ registers in their integer representation. This operation can be efficiently implemented in the Fourier domain as a concatenation of weighted addition operations~\cite{Ruiz2017} shown as $\Phi$ADD and controlled $\Phi$ADD blocks in Fig.~\ref{fig:multiplier}, preceded and followed by a QFT and IQFT respectively~\cite{Nielsen2010}. In the Grover-H variant the QFT and IQFT are not applied: the $Z$ register is initialised in $\ket{+}^{\otimes n_z}$ (the Hadamard basis), and the $\Phi$ADD gates act directly on this state without a preceding QFT. Since both the Fourier basis and the Hadamard basis are product states of equal-weight superpositions differing only in relative phases, the arithmetic phases accumulated by the $\Phi$ADD gates are identical in both cases; the QFT/IQFT pair is therefore unnecessary and can be omitted entirely. The circuit for $\Phi$ADD is depicted in Fig.~\ref{fig:adder}. It is built using a series of controlled phase shift gates with exponentially decreasing phase angles:

$$R^j(c) = \left[ {\begin{array}{cc}
   1 & 0 \\
   0 & e^{ic\pi/2^{j}} \\
  \end{array} } \right].
$$

Controlled $\Phi$ADD blocks used for multiplication can be implemented by replacing the controlled phase gates with doubly controlled phase gates. Doubly controlled phase gates can be subsequently decomposed to three controlled phase gates and two CNOT gates each~\cite{Barenco1995}.

\begin{sidewaysfigure}[ht]
\centering\begin{tikzpicture}
\node[scale=0.95] {
\begin{quantikz}
x_0~& &\ctrl{4}\gategroup[12, steps=4, style=dotted]{$+axy$} & &~\ldots~ &  & & \gate[8]{\mathrm{\Phi ADD}(b)} \gategroup[12, steps=1, style=dotted]{$+bx$}& &\gategroup[12, steps=1, style=dotted]{$+cy$} & & \gategroup[12, steps=1, style=dotted]{$+d$} &\\
x_1~& &  &  \ctrl{3 }& ~\ldots~ & & & & & & & &\\
\vdots \\
x_{n_x-1}~& &  &  &~ \ldots ~ & \ctrl{1} & & & & & & &\\
\lstick[wires=4]{$\phi(z)$} z_0~& &\gate[8]{\mathrm{\Phi ADD}(a)}&\gate[8]{\mathrm{\Phi ADD}(2a)}&~\ldots~ &\gate[8]{\mathrm{\Phi ADD}(2^{n_x}a)} & & & &\gate[8]{\mathrm{\Phi ADD}(c)}& & \gate{R^0(d)} & \rstick[wires=4]{$\phi(z+axy+bx+cy+d)$}\\
z_1~& &  &  & ~\ldots~ & & & & &&&\gate{R^1(d)}  &\\
\vdots \\
z_{n_z-1}~&  &  &  &~ \ldots ~ & && &&& &\gate{R^{n_z-1}(d)}  &\\
y_0~& & & &~\ldots~ & & &  & & & & &\\
y_1~& &  & & ~\ldots~ & & & & & & & &\\
\vdots \\
y_{n_y-1}~& &  &  &~ \ldots ~ & & &  & & & & &
\end{quantikz}
};
\end{tikzpicture}
\caption{$\mathrm{\Phi QMA(a)}$. Quantum circuit for multiplying and adding two registers with integer scaling constants $a, b, c, d$}\label{fig:multiplier}
\end{sidewaysfigure}

\begin{sidewaysfigure}[ht]
\begin{tikzpicture}
\node[scale=0.85]{
\begin{quantikz}[column sep = 0.2cm]
y_0~&\ctrl{4} &\ctrl{5} &\ctrl{6}& \ \ldots\ &\ctrl{8} & \ctrl{9} &  & & \ \ldots\ &  & & \ \ldots\ &  &  &\\
y_1~&  &  &  & \ \ldots\ & &  & \ctrl{4} & \ctrl{5} & \ \ldots\ &\ctrl{7} & \ctrl{8} & \ \ldots\ &  &  &\\
\ \vdots\ \\
y_{n_y-1}~&  &  &  & \ \ldots\ & &  &  &  & \ \ldots\ & & & \ \ldots\ & \ctrl{5} & \ctrl{6} &\\
\lstick[wires=6]{$\phi(z)$}
z_0~&\gate{R^0(c)} &  &  & \ \ldots\ & &  &  &  &\ \ldots\ &  &  & \ \ldots\ & &  & \rstick[wires=6]{$\phi(z+cy)$}\\
z_1~&  &\gate{R^1(c)} &  & \ \ldots\ & &  &\gate{R^0(c)} &  & \ \ldots\  & & & \ \ldots\ &  & &\\
z_2~& \qw &  &\gate{R^2(c)} & \ \ldots\ & &  &  &\gate{R^1(c)}& \colorbox{white!30}{\ldots}  &   \qw& & \ \ldots\ &  & &\\
\ \vdots\ \\
z_{n_z-2}~&  &  &  & \ \ldots\ & \gate{R^{n_z-2}(c)} &  &  &  & \ \ldots\ & \gate{R^{n_z-3}(c)} &  & \ \ldots\ & \gate{R^{n_z-n_y -1}(c)} &  &\\
z_{n_z-1}~&  &  &  &  \ \ldots\ & &\gate{R^{n_z-1}(c)} &  &  & \ \ldots\ & & \gate{R^{n_z-2}(c)}& \ \ldots\  &  & \gate{R^{n_z-n_y}(c)} &
\end{quantikz}
};
\end{tikzpicture}
\caption{$\mathrm{\Phi ADD(c)}$. Quantum circuit for adding two registers with integer scaling constant $c$}\label{fig:adder}
\end{sidewaysfigure}

\section*{S4. Decomposition of DIFF in terms of elementary gates}

The diffusion operators DIFF$_{X,Y}$ and DIFF$_{Z}$ realize the operation $DIFF_Q = I - 2\ket{s_Q}\bra{s_Q}$ where $\ket{s_Q}$ is the uniform state over the register Q,
$$\ket{s_Q} = \frac{1}{\sqrt{2^{n_q}}}\sum_{q=0}^{2^{n_q}-1}\ket{q}.$$
This can be implemented by the parallel application of H followed by X gates to each qubit, then a multicontrolled-Z gate, followed by another parallel application of X followed by H gates, as depicted in Fig.~\ref{fig:diff}. The multicontrolled-Z gate on $n_q$ qubits is implemented via phase kickback as $H \cdot C^{n_q-1}X \cdot H$ on the target qubit, where the $(n_q-1)$-fold multicontrolled-X is decomposed using the dirty-ancilla ladder of~\cite{Barenco1995} (Lemma 7.2) with the other register(s) supplying the $n_q - 3$ required dirty ancilla qubits, giving a cost of exactly $4(n_q - 3)$ Toffoli gates per DIFF call.

\begin{figure}[ht]
\centering
\begin{quantikz}[column sep = 0.15cm, row sep = 0.4cm]
\lstick{$q_0$} & \gate{H} & \gate{X} & \qw       & \qw       & \qw       & \ctrl{6}  & \qw       & \qw       & \qw       & \ctrl{6}  & \qw       & \qw       & \gate{X} & \gate{H} & \qw \\
\lstick{$q_1$} & \gate{H} & \gate{X} & \qw       & \qw       & \qw       & \ctrl{5}  & \qw       & \qw       & \qw       & \ctrl{5}  & \qw       & \qw       & \gate{X} & \gate{H} & \qw \\
\lstick{$q_2$} & \gate{H} & \gate{X} & \qw       & \qw       & \ctrl{3}  & \qw       & \ctrl{3}  & \qw       & \ctrl{3}  & \qw       & \ctrl{3}  & \qw       & \gate{X} & \gate{H} & \qw \\
\lstick{$q_3$} & \gate{H} & \gate{X} & \qw       & \ctrl{1}  & \qw       & \qw       & \qw       & \ctrl{1}  & \qw       & \qw       & \qw       & \qw       & \gate{X} & \gate{H} & \qw \\
\lstick{$q_4$} & \gate{H} & \gate{X} & \gate{H}  & \targ{}   & \qw       & \qw       & \qw       & \targ{}   & \qw       & \qw       & \qw       & \gate{H}  & \gate{X} & \gate{H} & \qw \\
\lstick{$q_5$} & \qw      & \qw      & \qw       & \ctrl{-1} & \targ{}   & \qw       & \targ{}   & \ctrl{-1} & \targ{}   & \qw       & \targ{}   & \qw       & \qw      & \qw      & \qw \\
\lstick{$q_6$} & \qw      & \qw      & \qw       & \qw       & \ctrl{-1} & \targ{}   & \ctrl{-1} & \qw       & \ctrl{-1} & \targ{}   & \ctrl{-1} & \qw       & \qw      & \qw      & \qw \\
\end{quantikz}
\vspace{10pt}
\caption{$\mathrm{DIFF}$. Quantum circuit implementing the diffusion operation on $n_q = 5$
qubits ($q_0$--$q_4$), with two dirty ancilla qubits ($q_5$, $q_6$) drawn from the other
register. The multicontrolled-Z on $q_0$--$q_4$ is realised via phase kickback
($H \cdot C^{4}X \cdot H$ on $q_4$), where $C^4X$ is decomposed using the
dirty-ancilla ladder of Ref.~\cite{Barenco1995} (Lemma~7.2), giving $4(n_q-3)$
Toffoli gates.}\label{fig:diff}
\end{figure}

Table~\ref{tab:gates} summarizes the number of elementary gates needed for implementing each operation.

\begin{table}[ht]
    \centering
    \begin{tabular}{l|ll}
        \hline
        Primitive & Calls & Two-qubit gate count \\
        \hline
        $\Phi$Q(N)MA    & $2K$ & $\frac{1}{2}(5n_xn_y+n_x+n_y)(2n_z-n_x-n_y) + \frac{1}{2}(12n_xn_y+n_x+n_y)$ \\
        DIFF$_{Z}$   & $K$  & $24(n_z - 3)$ \\
        DIFF$_{X,Y}$ & $K$  & $24(n_x + n_y - 3)$ \\
        \hline
    \end{tabular}
    \caption{Two-qubit gate counts for the Grover-H algorithm after decomposing
    CCP$(\lambda)$ into $3\,\mathrm{CP} + 2\,\mathrm{CX}$ and Toffoli into $6\,\mathrm{CX}$.
    The $\Phi$Q(N)MA count is an upper bound; zero-angle pruning reduces the CCP contribution
    by an $N$-dependent amount.
    The number of Grover iterations is $K = \lfloor(\pi/4)\sqrt{2^{n_x+n_y}/n_\mathrm{sol}}\rfloor$,
    where $n_\mathrm{sol} \in \{1,2\}$ is the number of solutions in the search space.}
    \label{tab:gates}
\end{table}

\section*{S5. Digital adiabatic factoring algorithm}

The oracle construction can be embedded in quantum algorithms beyond Grover search; we illustrate this with a digital adiabatic variant. In adiabatic quantum computing the solution to a computational problem is encoded in the ground state of a Hamiltonian $H_P$. Starting from the readily prepared ground state of an initial Hamiltonian $H_I$ one leverages the adiabatic principle to slowly transform the system to the ground state of $H_P$.

We consider the time-dependent Hamiltonian $H(k) = (1-k/K)H_I + (k/K)H_P$ realised on a digital quantum computer through the Trotter expansion. Up to order $\mathcal{O}(\epsilon^2)$ the time evolution operator for step $k$ is approximated as
\begin{equation}\label{eq:adiabatic}
U_k \approx e^{-i\epsilon(1-k/K)H_I}\times e^{-i\epsilon(k/K)H_P}.
\end{equation}
The initial state for $X$, $Y$ coincides with the ground state of $H_I = -\frac{1}{2}\sum_j\sigma_x^j$. $H_P$ is given by
\begin{align}\label{eq:HP}
H_P = QMA\left (\sum_{\zeta=0}^{n_z-1}P^\zeta_1\right )QNMA,
\end{align}
where $P_1^\zeta = \ket{1}_\zeta\bra{1}$, $QNMA = \sum_{x,y,z}\ket{x,y, z-f(x,y) \bmod 2^{n_z}}\bra{x, y, z}$, and $QMA = QNMA^\dagger$.

Algorithm~3 (below) presents the digital adiabatic factoring algorithm, constructed from Algorithm~1 of the main text by replacing the diffusion operator $\mathrm{DIFF}_{XY}$ with single-qubit rotation gates $\mathrm{RX}(2\beta_k) = \exp(-i\beta_k\sigma_x)$ and replacing the binary phase mark with a parametric multi-controlled phase $e^{i\gamma_k|+\rangle\langle+|_Z}$ acting on the $Z$ register. The angles follow a fixed adiabatic schedule $\gamma_k = (k/K)\pi$ and $\beta_k = (1-k/K)\pi/2$ for $k=1,\ldots,K$, without classical optimisation. We find this approach converges faster than implementations which directly minimise $H_P = (N-\hat p \hat q)^2$~\cite{Peng2008, Hegade2021}, which we attribute to the smaller spectral norm of Eq.~(\ref{eq:HP}), of $\mathcal{O}(poly(n))$ versus $\mathcal{O}(poly(2^n))$~\cite{Aharonov2008, Kieu2019A}. The adiabatic algorithm's convergence is not guaranteed (in case of a vanishing spectral gap) and it is an open question which problems allow a computational quantum speedup in adiabatic quantum computing~\cite{vanDam2001, Troels2014, Hastings2021}.

As shown in Fig.~2 of the main text, for $N = 101911 = 223\times 457$ the digital adiabatic algorithm requires many more steps than Grover-H, reaching $P>0.9$ only after $>10^4$ steps with a linear schedule. The Grover algorithm converges more deterministically, while the adiabatic algorithm requires between $2^{n-4}$ and $2^{n/2-1}$ steps with large scatter depending on the problem instance. In terms of two-qubit gates, the digital adiabatic algorithm is competitive with Grover for only $\lesssim 0.1$ of problem instances studied.

\begin{algorithm}[b]
\small
\caption{Factorize($N$) using digital adiabatic evolution}
\begin{algorithmic} \label{alg:3}
    \STATE \textbf{Input:}
    \bindent
        \STATE $\bullet$ The integer $N$ to be factored and the number of iteration steps $K$.
        \STATE $\bullet$ Three registers $X, Y, Z$ containing $n_x$, $n_y$ and $n_z = n_x+n_y+3$ qubits respectively.
    \eindent
    \STATE \textbf{Output:}
    \bindent
        \STATE $\bullet$ Two bitstrings $x=a, y=b$ satisfying $N = (6a+p_0)(6b+q_0)$
    \eindent
    \STATE \textbf{Procedure:}
    \bindent
        \STATE \textbf{Step 0.} Check that $N$ is not a multiple of $2$ or $3$. Otherwise return the trivial factors.
        \STATE \textbf{Step 1.} Initialize registers\\
        \qquad\qquad$\bullet$ Prepare $X$ and $Y$ registers in the uniform state $\frac{1}{\sqrt{2^{n_x+n_y}}}\left (\ket{0}+\ket{1}\right )^{\otimes (n_x+n_y)}$\\
        \qquad\qquad$\bullet$ Prepare $Z$ register in the Fourier basis via QFT$\ket{0}^{n_z}$
        \STATE \textbf{Step 2.} Choose $s=+1$; set $\gamma_k = (k/K)\pi$,\; $\beta_k = (1-k/K)\pi/2$
        \FOR{ $1 \leq k \leq K$}
        \bindent
        \STATE \textbf{Step 3.} Apply $H_P$ terms\\
        \qquad\qquad$\bullet$ Apply QNMA: $\ket{x,y,z}\rightarrow \ket{x,y,z-f(x,y)}$\\
        \qquad\qquad$\bullet$ Apply parametric phase $e^{i\gamma_k|{+}\rangle\langle{+}|_Z}$ on $Z$\\
        \qquad\qquad$\bullet$ Apply QMA: $\ket{x,y,z}\rightarrow \ket{x,y,z+f(x,y)}$
        \STATE \textbf{Step 4.} Apply transverse field $\bigotimes_j^{X,Y} \mathrm{RX}(2\beta_k)$
        \eindent
        \ENDFOR
    \STATE \textbf{Step 5.} Apply IQFT on $Z$. Measure the $X$, $Y$ registers. Calculate $p=6x+p_0,\; q=6y+q_0$.\\
    \qquad\qquad\quad\quad If $N\neq pq$ restart with $s=-1$ in Step 2.
    \eindent
\end{algorithmic}
\end{algorithm}

\section*{S6. Ross--Selinger synthesis for Grover-H}

Grover-H uses controlled-phase gates CP$(\theta)$ and CCP$(\theta)$ with rotation angles $\theta = 2\pi c / 2^{n_z}$ that are not multiples of $\pi/4$ when $c$ is not a power of two. These are approximated by Clifford+T sequences using the Ross--Selinger gridsynth algorithm~\cite{RossSelinger2016}, which finds the optimal-T-count sequence for a given precision $\varepsilon$, with T-count scaling as $\mathcal{O}(\log(1/\varepsilon))$.

We evaluated synthesis precision over 24 prime pairs with $6 \leq n \leq 14$, sweeping $\varepsilon$ from $0.5$ to $10^{-4}$ and recording the ratio $P_\mathrm{CT}/P_\mathrm{native}$ of success probability after Clifford+T transpilation to the native-gate success probability. For all instances with $n \leq 13$, a precision $\varepsilon \leq 0.1$ achieves $P_\mathrm{CT}/P_\mathrm{native} \geq 0.9$. For the largest instances ($n = 14$, $N = 3127$--$8453$) a tighter precision $\varepsilon \lesssim 0.007$ is required. The Grover-AND variant requires no synthesis and has an exact T-gate count of $7 n_\mathrm{CCX}$ per oracle call.

\bibliographystyle{iopart-num}
\bibliography{main}